\documentclass{aa}  
\usepackage{graphicx, xcolor}
\usepackage{txfonts}
\usepackage{amsmath}
\usepackage{float}
\usepackage{dblfloatfix}
\bibpunct{(}{)}{;}{a}{}{,}
\usepackage{natbib,twoopt}
\usepackage{bm}
\usepackage{url}
\usepackage[hidelinks]{hyperref}
\usepackage{array}
\usepackage{booktabs}
\usepackage{xfrac}
\usepackage{subcaption}
\usepackage{comment}
\hypersetup{
    colorlinks=true,
    linkcolor=blue,
    filecolor=magenta,
    citecolor=blue,
    urlcolor=blue,
    breaklinks=true
    }

\usepackage{soul}
\usepackage{lipsum}
\usepackage{subcaption}         % necessary for continued figures, example in section 3
\usepackage{lscape}             % to rotate a single page table, example in appendix.
\usepackage{placeins}           % useful with \FloatBarrier, to keep 
\begin{document}

%%%%%%%%%%%%%%%%%%%%%%%%%%%%%%%%%%%%%%%%
% if you use custom commands in your title,
% ensure to check your title when submitting!
%%%%%%%%%%%%%%%%%%%%%%%%%%%%%%%%%%%%%%%%
   \title{Improved delensing of CMB $B$-modes with upcoming SO and LSST Year-1 mock dataset}
   \subtitle{Impact of photometric redshift errors}

%%%%%%%%%%%%%%%%%%%%%%%%%%%%%%%%%%%%%%%%
% Please separate each author with the \and command
%
% Use the \corrauth to provide the corresponding
% author address. It will be automatically inserted as 
% footnote in the PDF output.
%
% Please DO NOT include ORCIDs next to author names.
% Instead, please provide an active address for each coauthor:
% it will be automatically extracted by EDPS editorial system, 
% and co-authors will be be able to authenticate their ORCID.
%
% Only authenticated ORCIDs will be taken into account.
% ORCIDs included here will be removed.
%%%%%%%%%%%%%%%%%%%%%%%%%%%%%%%%%%%%%%%%

   \author{ Kishan Deka\inst{1}\thanks{Corresponding author:         \href{mailto:kishan.deka@ncbj.gov.pl}{\texttt{kishan.deka@ncbj.gov.pl}}}       % use \corrauth for the corresponding author
        \and Chandra Shekhar Saraf\inst{2}
        %\fnmsep\thanks{NASA fellow (shows the usage of elements in the author field)}
        \and Pawe{\l} Bielewicz\inst{1}
        }

   \institute{National Centre for Nuclear Research, Pasteura 7, 02-093, Warsaw, Poland
   \and Korea Astronomy and Space Science Institute, 776 Daedeok-daero, Yuseong-gu, Daejeon 34055, Republic of Korea}

   \date{Received October 08, 2026}

% \abstract{}{}{}{}{}
% 5 {} token are mandatory
 
  \abstract{In the era of high-precision Cosmic Microwave Background (CMB) anisotropy measurements, detection of  primordial gravitational waves (PGWs) is one of the main goals of current CMB experiments. Detection of a divergence-free ($B$-mode) component in CMB polarisation will provide direct evidence of inflationary PGWs in the early universe. However, weak gravitational lensing of CMB photons due to intervening mass distribution of the Universe distorts the primordial polarisation patterns, introducing an additional lensing-induced $B$-mode signal. Removing these lensing-induced $B$-modes is essential for improving constraints on the amplitude of PGWs. In this work, we present a multi-tracer delensing analysis using simulations of the Rubin Observatory LSST galaxy catalogue and Simons Observatory (SO) CMB maps. We use realistic photometric redshift errors for LSST mock catalogue and high complexity Galactic foregrounds for CMB simulations. By adopting a tomographic multi-tracer approach that combines the CMB and LSST mock data, we achieved a 10-20\% higher delensing efficiency compared with using CMB data alone. Our analysis removes approximately 60\% of the lensing-induced $B$-mode power. We find that photometric redshift errors in the mock galaxy catalogue introduce biases in the inferred galaxy bias and $\sigma_{8,0}$ parameters. However, these uncertainties do not produce a significant systematic bias in the inferred tensor-to-scalar ratio.
}
   
    \titlerunning{Multi-tracer delensing of CMB $B$-modes with upcoming SO+LSST-Y1 data}
    \authorrunning{Deka, K., Saraf, C.S. \& Bielewicz, P.}              
    \maketitle

%%%%%%%%%%%%%%%%%%%%%%%%%%%%%%%%%%%%%%%%%%%%%%%%%%%%%%%%%%%%%%
\section{Introduction}
The standard model of cosmology, characterised by six phenomenological parameters, assumes a near-exponential expansion during the first fraction of a second after beginning of the Universe. This period, known as inflation, explains the spatial flatness and large-scale isotropy of our Universe \citep{Starobinskii_1979, Guth_1982, Linde_1983}. According to inflationary models, primordial scalar and tensor quantum fluctuations have been amplified by the rapid expansion of the Universe. The scalar fluctuations acted as a primary seed for structure formation at later times, while the tensor fluctuations may have produced a stochastic primordial gravitational wave (PGW) background. The detection of such a PGW signal will be direct proof of inflation, and the amplitude of the PGW background will probe the energy scale of inflation, providing a window into the physics of the early Universe. 

The cosmic microwave background (CMB) radiation has been used as a direct observable to infer the cosmological parameters of the standard model \citep{WMAP_params_2013, Planck_VI_2020, ACT_params_2025}. The temperature and polarisation anisotropies of the CMB contain imprints of primordial fluctuations, which have been detected with several CMB experiments. More importantly, tensor fluctuations from inflationary PGW may have left their imprint as a parity-odd component in the CMB polarisation field \citep{Polnarev_1985, Kamionkowski_1997, Seljak_1997}. This primordial component, known as tensor B-mode due to its characteristic divergence-free pattern, provides a unique probe of PGW because it has no primary contribution from the scalar perturbations. At leading order, the scalar fluctuations only imprint a curl-free parity-even component, called E-mode, in the CMB polarisation. The ratio of the amplitude of the tensor fluctuations to scalar fluctuations is characterised by the tensor-to-scalar ratio parameter, $r$. Although the primordial tensor B-mode signal is yet to be detected, an upper limit on the tensor-to-scalar ratio $r < 0.032$ at $0.05$ Mpc$^{-1}$ (95\% confidence level) pivot scale exists from combined datasets \citep{Tristram_2022}. Ongoing CMB experiments such as Simons Observatory (SO)  aim to achieve a sensitivity of $\sigma(r=0) \lesssim 0.003$ by the end of the initial five-year survey \citep{SO_2019}.

For upcoming CMB data from high-sensitivity measurements, contamination of B-mode polarisation from secondary sources will be a major concern for detecting the primordial B-mode signal. There are two main sources of contamination in the B-mode polarisation at relevant scales -- one is the polarised emission from our Galactic medium \citep{Planck_XI_2020}, and the other is the secondary B-mode signal produced by weak lensing of CMB photons by the intervening mass distribution \citep{Zaldarriaga_1998}. The deflection of CMB photons by the large-scale structure (LSS) remaps the CMB temperature and polarisation anisotropies \citep{Lewis_2006}. Gravitational lensing distorts the primordial polarisation patterns, causing a leakage from primordial E-mode to B-mode patterns, which induce a secondary lensing B-mode signal in the observed CMB polarisation anisotropy maps \citep{Lewis_2001, Kesden_2002}. {After the initial five-year survey of SO, the noise level in B-mode polarisation is expected to achieve 2$\mu$K-arcminutes at large scales \citep{SO_2019}. At this noise level, the lensing B-mode signal will dominate the observed B modes.} To reduce the additional variance coming from lensing B modes, a delensing procedure is performed, where a template of lensing B-mode, constructed from the observed E-mode and the lensing convergence field, is subtracted from the observed B-mode map \citep{Seljak_2004, Smith_2012, Carron_2019, Belkner_2024}. The lensing B-mode template can be built in two ways -- (a) using a internally reconstructed lensing convergence field employing lensing induced non-Gaussian CMB anisotropies; (b) using an external LSS tracer as a proxy for the lensing convergence field.

Delensing on CMB temperature and polarisation data using internal CMB lensing reconstruction has been performed with Planck observations to achieve around 25\% of delensing efficiency \citep{Carron_2017}. Moreover, several recent studies have demonstrated the robustness of using external tracers in delensing to improve the delensing efficiency and constraints on tensor-to-scalar ratio \citep{Manzotti_2017, Planck_VIII_2018, BICEP_SPTpol_2021, Hertig_2024}. The background radiation originating from unresolved dusty star-forming galaxies with a redshift distribution peak at $z\sim2-3$, known as the cosmic infrared background (CIB), has been shown to be an useful external LSS tracer in delensing, achieving $\sim 28\%$ delensing efficiency \citep{Sherwin_2015, Larsen_2016}. Despite the limited depth of existing galaxy surveys in redshift, recent studies achieved $\sim 40\%$ delensing efficiency by combining the galaxy overdensities with other LSS tracer \cite{Yu_2017, Hertig_2025}. In upcoming years, the relevance of galaxy overdensities as a tracer in delensing will be even higher, due to dataset from Vera C. Rubin Observatory's Legacy Survey of Space and Time (LSST) collaboration. \cite{Manzotti_2017} has showed that a tomographic approach where the galaxy overdensity maps split into redshift bins are used as LSS tracers, it could achieve better delensing efficiency because of its increased correlation with the lensing kernel. 

In this work, we mainly focused on optimal removal of the lensing-induced secondary B-mode signal. We explored the tomographic multi-tracer delensing approach in an experimental setup considering SO CMB observations and galaxy catalogue from Rubin Observatory LSST survey.

This paper is structured as following. In Sect. \ref{sec:theory}, we present the mathematical framework of CMB weak lensing and LSS tracers. We discuss the methods on template-based delensing and estimation of delensing performance in Sect. \ref{sec:theory}. We present the methodology of covariance matrix estimation and likelihood analyses in Sec. \ref{sec:methods}. In Sec. \ref{sec:sims}, we discuss how the simulations for CMB and galaxy observations are obtained. In Sect. \ref{sec:results}, we present main results of the study and we conclude our work in Sect. \ref{sec:summary}.

%%%%%%%%%%%%%%%%%%%%%%%%%%%%%%%%%%%%%%%%%%%%%%%%%%%%%%%%%%%%%%
\section{Theoretical background}
\label{sec:theory}

\subsection{Weak lensing of CMB}
The CMB photons travelling from the last scattering surface to us experience gravitational lensing along its path by the intervening matter, which distorts the observed CMB anisotropies. We observe this lensing effect in CMB as a direction-dependent remapping of primordial CMB anisotropies, given as,
\begin{equation}
\begin{split}
    T(\hat{\mathbf{n}}) &= \widetilde{T}\left(\hat{\mathbf{n}}+\nabla \phi(\hat{\mathbf{n}})\right) \\
    [Q+iU](\hat{\mathbf{n}}) &= [\widetilde{Q}+i\widetilde{U}] \left(\hat{\mathbf{n}}+\nabla \phi(\hat{\mathbf{n}})\right)  \label{eq:deflection}
\end{split}
\end{equation}
Here, $T,\,Q,\,U$ denotes the Stokes parameters of lensed CMB temperature anisotropies and polarisation fields. Meanwhile, $\widetilde{T}$, $\widetilde{Q}$, $\widetilde{U}$ represent the CMB fields before lensing. The lensing-potential field, $\phi(\hat{\mathbf{n}})$, is defined as \citep{Hu_2002, Lewis_2006},
\begin{equation}
    \phi(\hat{\mathbf{n}}) = -2\int_{0}^{\chi^{*}} d\chi\, \frac{D_A(\chi^{*}-\chi)} {D_A(\chi)\,D_A(\chi^{*})} 
    \Psi(\chi(\hat{\mathbf{n}}),\eta)   \label{eq:lens_pot}
\end{equation}
where $\Psi$ denotes the Weyl gravitational potential evaluated along the photon path, and the integral extends over the comoving distance $\chi$ from the observer to the last-scattering surface at $\chi^{*}\simeq 14000\,\mathrm{Mpc}$. The conformal time at which the photon was located at distance $\chi$ in the direction $\hat{\mathbf{n}}$ is denoted by $\eta$. The function $D_A(\chi)$ represents the angular-diameter distance, which depends on the spatial curvature of the Universe. In the special case of a spatially flat Universe, this distance reduces to $D_A(\chi)=\chi$.

Using the spin-weight spherical harmonic formalism \citep{Goldberg_1967, Lewis_2001}, any spin-$s$ field ${}_sX(\hat{\mathbf{n}})$ on the sky can be expressed as,
\[
{}_sX(\hat{\mathbf{n}}) = \sum_{\ell m} {}_sX_{\ell m}\,{}_sY_{\ell m}(\hat{\mathbf{n}}) \;.
\]
where $\ell$ and $m$ represent the spherical harmonic multipole and azimuthal index for field $X$, respectively. The spin $s=0$ represents CMB temperature field ($_{0}X \equiv T$) and spin $s=2$ represents the CMB polarisation fields defined as $_{\pm2} X \equiv \, _{\pm2}P = Q \pm i\, U$. Then, the lensing-induced change in temperature and polarisation is obtained by applying the spin-weighted harmonic transforms to Eqs.~\eqref{eq:deflection} and \eqref{eq:lens_pot}. After integrating by parts and using the known analytic expression for the triple integral of spin-weighted spherical harmonics, one obtains
\begin{equation}
\begin{split}
    \delta_sX_{\ell m} & = {}_s\widetilde{X}_{\ell m}-{}_sX_{\ell m} \\
                    & =  \sum_{LM\ell'm'} \frac{(-1)^m}{L(L+1)}
                            \begin{pmatrix}
                            \ell & \ell' & L \\
                            m & -m' & -M
                            \end{pmatrix}
                                F^{(s)}_{\ell \ell' L} \,{}_sX_{\ell'm'}\,\kappa_{LM} \;, \label{eq:lensing_defl}
\end{split}
\end{equation}
Here, the quantity in round brackets denotes the Wigner-3$j$ matrix connecting multipole and azimuthal index between different fields. The lensing induced mode-coupling function $F^{(s)}_{\ell\ell'L}$ is defined in \citet{Okamoto_2003} and $\kappa$ denotes the lensing convergence which is defined in real space as,
\[ \kappa = -\frac{1}{2}\nabla^2\phi \;,\]
which can be written in terms of the harmonic coefficients as,
\[ \phi_{LM} = \frac{2}{L(L+1)}\kappa_{LM} \;.\]

Following \cite{Zaldarriaga_1997}, we can decompose the spin-2 polarisation fields in two scalar components : the parity-even $E$-mode component and the parity-odd $B$-mode component as, 
\[ E_{\ell m} = \frac{1}{2} \left({}_{+2}P_{\ell m}+{}_{-2}P_{\ell m}\right),
\text{ and } B_{\ell m} = \frac{1}{2i} \left({}_{+2}P_{\ell m}-{}_{-2}P_{\ell m}\right) . \]

The lensing contribution to the $B$-mode anisotropies follows directly from Eq. \eqref{eq:lensing_defl}, giving
\begin{equation}
B^{\mathrm{len}}_{\ell_1 m_1} = \sum_{\ell_2 m_2 \ell m}
        \begin{pmatrix}
        \ell_1 & \ell_2 & \ell \\
        m_1 & m_2 & m
        \end{pmatrix} \frac{g^{EB}_{\ell_1 \ell_2 \ell}}{\ell(\ell + 1)} E_{\ell_2 m_2}\,\kappa_{\ell m} \;,  \label{eq:lensing_Bmode}
\end{equation}
which represents a convolution between the unlensed $E$ modes and the lensing convergence, $\kappa$. The coupling coefficients, $g^{EB}_{\ell_1 \ell_2 \ell}$, are defined as \citep{Smith_2012},
\begin{equation}
g^{EB}_{\ell_1 \ell_2 \ell} = i\left(F^{(-2)}_{\ell_1 \ell_2 \ell} - F^{(2)}_{\ell_1 \ell_2 \ell}\right)  \;.
\end{equation}
These lensing-generated $B$ modes increase the cosmic variance and reduce the precision of constraints on primordial gravitational waves. To reduce this effect, a first-order estimate of the lensing $B$-mode pattern is constructed from observational data and then used for delensing.

\subsection{LSS tracers}
The underlying dark matter distribution of the Universe that governs the formation of LSS can not be probed directly. However several observables, including galaxy overdensity maps, CMB and galaxy weak lensing, and CIB maps, serve as tracers of the LSS. As the lensing potential field expressed in Eq. \eqref{eq:lens_pot}, the CIB and galaxy overdensity maps are also projected two-dimensional fields of three-dimensional distribution of matter in the Universe. The projected field for any tracer can then be expressed as a line-of-sight integral of the matter overdensity, $\delta_i$,
\begin{equation}
X^i(\hat{\mathbf{n}}) = \int_0^\infty dz\,W_X^i(z)\, \delta_i\left(\chi(z)\,\hat{\mathbf{n}},z\right) \;.
\end{equation} 
Here, $\chi(z)$ denotes the comoving distance corresponding to redshift $z$, while $W_X^i(z)$ specifies the projection kernel, which weights the contribution of matter fluctuations at each redshift for the field $X^i$. Using the Limber approximation \citep{Limber_1953}, the angular power spectrum between two LSS tracers, labelled by $i$ and $j$, can be written as
\begin{equation}
C^{ij}_{\ell} = \int_0^{z_*} dz\, \frac{H(z)}{c\,\chi^2(z)} W^i(z)\,W^j(z)\, P\left(k=\frac{\ell + 1/2}{\chi(z)},z\right) \;.
\end{equation}
Here, $H(z)$ is the Hubble parameter at redshift $z$, $c$ is the speed of light, and $P(k,z)$ denotes the matter power spectrum evaluated at redshift $z$. The kernels $W^i(z)$ and $W^j(z)$ describe the redshift-dependent contributions of the two tracers considered in this work. The integral covers the redshift range from us to the last scattering surface at redshift $z_* \simeq 1100$. 

The lensing convergence kernel in flat cosmology is given as,
\begin{equation} W^{\kappa}(z) = \frac{3\Omega_m}{2c}
\frac{H_0^2}{H(z)} (1+z)\chi(z) \frac{\chi_*-\chi(z)}{\chi_*} \;,
\end{equation}
where $\chi_*$ is the comoving distance to the last-scattering surface. The quantities $\Omega_m$ and $H_0$ denote the present-day matter density parameter and Hubble constant, respectively. The lensing kernel peaks around redshift $z\sim 2-3$ where most of the lensing sources lie and extends to high redshift.

The kernel for galaxy overdensities is written in terms of galaxy redshift distribution as,
\begin{equation}
    W^g(z) = b(z)\, \frac{dN}{dz} \;.
    \label{eq:galaxy_density_kernel}
\end{equation}
Here, $\frac{dN}{dz}$ is the redshift distribution of number of galaxies observed as a function of redshift which is normalised for each redshift bin. The redshift dependent galaxy bias $b(z)$ connects the amplitude of galaxy clustering to the dark matter halo distribution.

\subsection{Lensing reconstruction}
The lensing convergence field traces the matter distribution up to very high redshift, but the lensing field can not be probed directly. Instead, an estimate of the lensing convergence field is reconstructed from a combination of observed lensed CMB maps, using the well-known quadratic estimator (QE) \citep{Hu_2002, Okamoto_2003, Planck_VIII_2018}. Following the full-sky formalism presented in \citet{Okamoto_2003}, we can estimate the lensing convergence field as,
\begin{equation}
    \hat{\kappa}_{LM}^{XY} = \frac{A_L^{XY}}{\left(L(L+1)\right)^2} \sum_{\ell m}\sum_{\ell^\prime m^\prime} (-1)^M \begin{pmatrix}
         \ell & \ell^\prime & L \\
         m & m^\prime & -M 
    \end{pmatrix}
    \mathcal{W}^{XY}_{\ell L \ell'} ~ \overline{X}_{\ell m} \, \overline{Y}_{\ell'm'} \; .  \label{eq:lens_qest}
\end{equation}
Here, $\overline{X}_{\ell m}$ and $\overline{Y}_{\ell'm'}$ are inverse variance-filtered observed CMB maps given as,
    \begin{equation} \label{eq:inv_weight}
        \overline{X}_{\ell m} = \left(C^{XX}_\ell+N^{\rm res}_\ell + F^{\rm res}_\ell\right)^{-1} \hat{X}_{\ell m} \; , 
    \end{equation}
where $N_{\ell}^{\rm res}$ and $F_{\ell}^{\rm res}$ are the power spectra for residual instrumental noise and residual Galactic foreground present in component separated maps, respectively. The normalisation factor, $A_L^{XY}$, and the minimum variance weights, $\mathcal{W}^{XY}_{\ell \ell^\prime L}$, are defined in \citet{Okamoto_2003}.  The reconstructed lensing convergence contains the reconstruction noise which depends on instrumental noise levels and the resolution of the experiment. The angular power spectra of the reconstructed lensing field is given as,
\begin{equation}
    \left< \hat{\kappa}^{XY} \hat{\kappa}
    ^{X'Y'} \right> =  C_L^{\kappa\kappa} + N_L^{\kappa\kappa} \;.
\end{equation}
The lensing reconstruction noise is computed analytically following the formalism presented in \cite{Deka_2026}. Multiple combinations of CMB fields can be used in Eq. \eqref{eq:lens_qest} by choosing different pairs of $X,Y$ fields, such as $TT, TE, EE, EB, TB$. In this work, we implement a minimum variance combination of all five pairs \citep{Okamoto_2003}. By constructions, these kinds of minimum variance estimators have the least level of reconstruction noise. Following \citet{Deka_2026}, we estimated the mean field contribution using 200 realisations of CMB observations with SO's instrumental specification and sky coverage, and subtracted the mean field from our estimate.

\subsection{Delensing of $B$-modes}
We subtract the lensing template from observed $B$ modes to obtain the delensed $B$-mode,
\begin{equation}
    B_{\ell m}^{\rm del} = B^{\rm obs}_{\ell m} - B^{\rm temp}_{lm}
\end{equation}
where, $B^{\rm obs}_{lm}$ is the observed $B$-modes which contains instrumental noise and Galactic foreground contamination, and $B^{\rm temp}$ is a template of lensing $B$-modes obtained by taking a convolution between observed $E$ modes and a tracer of LSS.
\begin{equation}
\begin{split}
    B^{\rm temp}_{\ell_1 m_1} = \sum_{\ell_2 m_2 \ell m} \mathcal{W}_{\ell_1 \ell_2 \ell}^{EB} & \begin{pmatrix}
        \ell_1 & \ell_2 & \ell \\
        m_1 & m_2 & m
    \end{pmatrix}  \times \\
    & \left( \frac{C^{EE}_{\ell_2}}{C^{EE}_{\ell_2} + N^{EE}_{\ell_2}} \right)
     \left( \frac{C^{\kappa I}_{\ell}}{C^{II}_{\ell} + N^{II}_{\ell}} \right)  E^{\rm obs}_{\ell_2 m_2}\, I_{\ell m} \;.
\end{split}  \label{eq:lensing_Btemp}
\end{equation}
Here, the observed $E$ modes, $E^{\rm obs}_{\ell m}$, and the tracer of convergence field, $I_{\ell m}$, are Wiener filtered with a factor $C_\ell \big/ \big(C_\ell + N_\ell \big)$, where $N_\ell$ is the noise corresponding to the measurement of these fields. The Wiener factor gives a optimal weighting by minimising the variance of the delensed $B$-mode map. Then, subtracting this template from the lensed $B$ modes, one can obtain the residual lensing power spectra as,
\begin{equation}
    C_{\ell_1}^{BB,{\rm res}} = \frac{1}{2\ell_1+1} \sum_{\ell_2\,\ell} 
    \left|\mathcal{W}^{EB}_{\ell_1\ell_2\ell}\right|^2 
    \left[ C_{\ell_2}^{EE}\,C_{\ell}^{\phi\phi} - 
    \left( \frac{\left(C_{\ell_2}^{EE}\right)^2} {C_{\ell_2}^{EE}+
    N_{\ell_2}^{EE}} \right) \rho_{\ell}^2 \right] \;,
\label{eq:residual_delensing}
\end{equation}
with correlation coefficient
\begin{equation}
    \rho_{\ell}^2 = \frac{\left(C_{\ell}^{\kappa I}\right)^2} 
    {(C_{\ell}^{II}+N_{\ell}^{II})\, C_{\ell}^{\kappa\kappa}} \;. \label{eq:rho_l_cmb}
\end{equation}

The internal delensing method is where we use the reconstructed lensing convergence field itself as the LSS tracer in Eq. \eqref{eq:lensing_Btemp}, then the correlation coefficient simplifies as,
\begin{equation}
    \rho_{\ell}^2 = \frac{\left(C_{\ell}^{\kappa\hat{\kappa}}\right)^2}{(C^{\kappa\kappa}_{\ell}+N^{\kappa\kappa}_{\ell})\,C^{\kappa\kappa}_{\ell}} = \frac{C^{\kappa\kappa}_{\ell}}{C^{\kappa\kappa}_{\ell}+N^{\kappa\kappa}_{\ell}}
    \label{eq:rho_internal}
\end{equation}
This method is internal in a sense that all the information are derived from CMB observations only to perform delensing.

\subsubsection{Delensing performance}
The angular power spectra of a delensed B-mode map is approximately given as,
\begin{equation}
\begin{split}
    C^{BB,\rm del}_{\ell} & \approx  C_{\ell}^{BB , \rm obs} - C_{\ell}^{BB,\rm temp}~ \\
                    & = r~C_{\ell}^{BB,\rm tens}(r=1) + C_{\ell}^{BB,\rm res} +  N_{\ell}^{BB} \\
\end{split}
\label{eq:delen_ps}
\end{equation}
where, 
\[
C_{\ell}^{BB,\rm res} = C_{\ell}^{BB,\rm lens} - C_{\ell}^{BB,\rm temp}. 
\]
Here, $C_{\ell}^{BB,~\mathrm{tens}}(r=1, n_t=0, n_{t,\rm run}=0)$ is the primordial tensor $B$-mode spectrum for $r=1$, $C_{\ell}^{BB, \mathrm{res}}$ is the residual lensing $B$-mode power after delensing as described in Eq. \eqref{eq:residual_delensing}, $N_{\ell}^{BB}$ is the instrumental noise power spectrum. 

If the $EB$ estimator is present in the lensing reconstruction, there could be a delensing bias produced due to correlated signal between the $B$-mode map used in the quadratic estimator and in delensing. But this contribution could be avoided using only higher order multipoles, i.e. $\ell > 300$, in the $EB$ quadratic estimator since we are only interested in large-scale $B$-modes for $r$ constraint to avoid any internal correlation. For the delensed large-scale $B$ modes, we can also define a multipole averaged lensing residue parameter as,
\begin{equation}
    A_{\rm lens} = 1 - \left<\frac{C_{\ell}^{BB, \rm temp}}{C_{\ell}^{BB, \rm lens}} \right>\Bigg|_{50 \leq \ell \leq 300}
\end{equation}
The $A_{\rm lens}$ parameter is a scaling parameter to quantify how much lensing signal is remaining in the delensed $B$-mode map. We can fit the $A_{\rm lens}$ parameter using a model for the delensed power spectra given as,
\begin{equation}
    C_{\ell}^{BB, \rm del} =  r~C_{\ell}^{BB,\rm tens}(r=1) + A_{\rm lens }\;C_{\ell}^{BB,\rm lens} +  N_{\ell}^{BB} 
    \label{eq:delensed_Bmode_model}
\end{equation}

\subsubsection{Multi-tracer approach}
Following \citet{Manzotti_2017}, we extend the delensing formalism to the multi-tracer approach where the tracers are distributed in redshift space. We consider a galaxy catalogue divided into redshift bins to give $n$ LSS tracers which can be combined optimally \citep{Sherwin_2015},
\begin{equation}
    I = \sum_{i=1}^n c_i I^i,  \label{eq:optimal_tracer}
\end{equation}
where
\begin{equation}
    c_i = [C_{II}^{-1}]_{ij}\, C^{\kappa I^j}. \label{eq:optimal_weights}
\end{equation}
Here, $C^{-1}_{II}$ is the inverse covariance matrix consists of the auto and cross-spectra between the $n$ number of tracers, i.e. $[C_{II}]_{ij} = C_{\ell}^{I^iI^j}$. The auto power spectra in this covariance include the noise term. For galaxy auto spectra, it is the shot noise white spectra, while for lensing convergence power spectra the noise term is lensing reconstruction noise. For multiple tracers, we can generalise the delensing efficiency from Eq. \eqref{eq:rho_l_cmb}  as,
\begin{equation}
    \rho_{\ell}^2 = \sum_{i,j} \frac{C_{\ell}^{\kappa I^i}\,  [C_{II}^{-1}]_{ij}\, C_{\ell}^{\kappa I^j}}{C_{\ell}^{\kappa\kappa}} \label{eq:rho_optimal}
\end{equation}

Splitting the galaxy catalogue in tomographic bins provides a higher correlation factor, which improves the delensing efficiency \citep{Manzotti_2017}. In case of photometric surveys, large errors in redshifts introduce cross-correlations between neighbouring bins. Hence, it becomes necessary to accurately account for all cross-bin correlations to avoid sub-optimal combinations of the tracers, which may affect the delensing efficiency.

\section{Simulations}
\label{sec:sims}
We used \texttt{CAMB}\footnote{\url{https://camb.readthedocs.io/}}\citep{2000ApJ...538..473L} to compute theoretical unlensed CMB TT, TE, EE power spectra and the matter power spectrum $P(k,z)$, assuming \textit{Planck} 2018 $\Lambda$CDM cosmology parameters \citep{Planck_VI_2020, Tristram_2024} and \texttt{HALOFIT} prescription \citep{2003MNRAS.341.1311S} for non-linear evolution. We used \texttt{GLASS}\footnote{\url{https://glass.readthedocs.io/stable/index.html}} \citep{2023OJAp....6E..11T} to generate correlated mock LSST galaxy catalogue and Simons Observatory CMB lensing convergence map. The methodology for mock catalogue and maps creation is detailed in the following sections.

\begin{table*}
\centering
\caption{
Noise and beam specifications used to produce the simulations LAT and SAT polarisation maps.
}
\begin{tabular}{ccccccc}
\hline\hline
\rule{0pt}{2.6ex} & \multicolumn{4}{c}{SAT ($f_{\rm sky} = 0.1$)} & \multicolumn{2}{c}{LAT ($f_{\rm sky} = 0.4$)} \\ 
\hline
\rule{0pt}{2.6ex}  Frequency [GHz] 
& FWHM [arcmin] 
& $\sigma_{\rm white}$ [$\mu{\rm K}$-arcmin] 
& $\ell_{\rm knee}$ 
& $\alpha_{\rm knee}$
& FWHM [arcmin] 
& $\sigma_{\rm white}$ [$\mu{\rm K}$-arcmin]  \\
\hline \rule{0pt}{2.6ex} 
27  & 91 & 33  & 15 & $-2.4$ & 7.4 & 74 \\
39  & 63 & 22  & 15 & $-2.4$ & 5.1 & 38 \\
93  & 30 & 2.5 & 25 & $-2.5$ & 2.2 & 8.2 \\
145 & 17 & 2.8 & 25 & $-3.0$ & 1.4 & 8.9 \\
225 & 11 & 5.5 & 35 & $-3.0$ & 1.0 & 21 \\
280 & 9  & 14  & 40 & $-3.0$ & 0.9 & 52 \\
\hline
\end{tabular}
\tablefoot{The white-noise levels are given for the central region of the map assuming five years of observations. The white-noise level of LAT temperature maps are $\sqrt{2}$ times smaller than the polarisation white-noise level.}
\label{tab:instrument_specs}
\end{table*}

\subsection{LSST galaxy catalogue}\label{sec:lsst_simulations}

We generated mock LSST catalogue with sky coverage area of $\sim 16,000 \text{ deg}^{2}$ and galaxy number density of 30 gal arcmin$^{-2}$, consistent with Y1 of LSST observations. We used \texttt{RAIL}\footnote{\url{https://github.com/LSSTDESC/RAIL}} \citep{2025arXiv250502928T} to sample LSST Y1 magnitudes of galaxies and their errors in $ugrizy-$bands. We estimated photometric redshifts of galaxies using the \texttt{FlexZBoost} estimator \citep{2017arXiv170408095I} built inside RAIL. To estimate these photometric redshifts, we created a training set and test set of $3\times 10^{5}$ and $10^{5}$ galaxies, respectively. With this setup, \texttt{FlexZBoost} yields a point-estimate of every galaxy photometric redshifts and their posteriors. We bin the galaxies into eight equal density tomographic bins based on their photometric redshifts point estimates. The fiducial values of galaxy bias in the eight tomographic bins is given as $b_{g} = [1.05,1.10,1.18,1.26,1.35,1.45,1.70,2.0]$.

\begin{figure}[hbt!]
    \centering
    \includegraphics[width=0.8\linewidth]{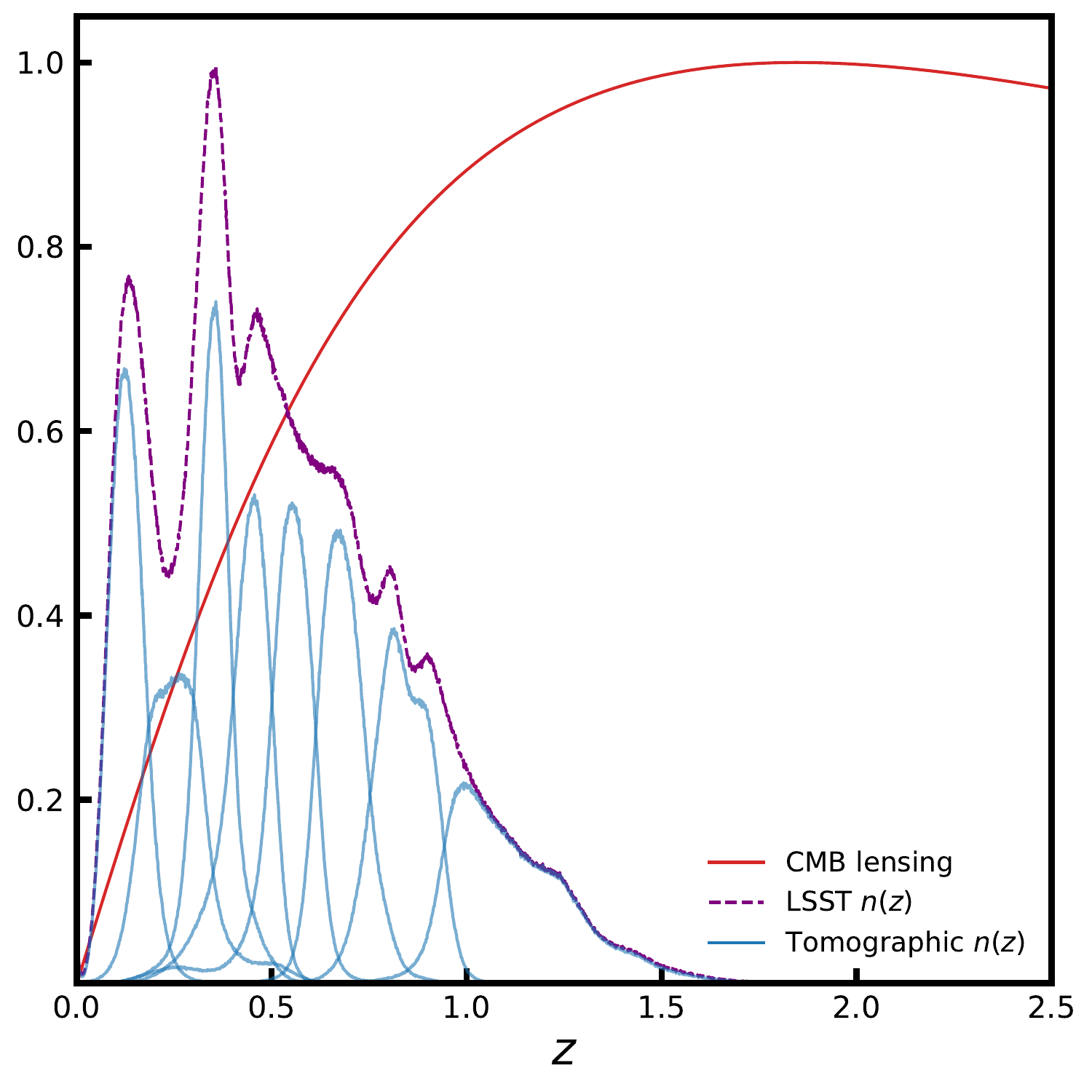}
    \caption{Tomographic LSST photometric redshift distributions from simulated catalogue (shown with blue lines) compared to CMB lensing kernel (red solid line). The dashed purple line shows the full redshift distribution of simulated LSST galaxies. Both full LSST redshift distribution and CMB lensing kernel have been rescaled to a unit maximum, for illustrative purposes only.}
    \label{fig:kernels}
\end{figure}

We stack the photometric redshift posteriors of galaxies to emulate the redshift distributions of the tomographic bins required for the galaxy density kernel given in Eq.\,\ref{eq:galaxy_density_kernel}. The redshift distributions of LSST galaxy samples divided into eight tomographic bins are shown in Fig. \ref{fig:kernels}. We also simulated a CMB lensing convergence map correlated with LSST galaxies, which was used to generate lensed CMB temperature and polarisation maps consistent with SO as described in Section \ref{sec:simons_maps}.

\subsection{Simons Observatory maps}\label{sec:simons_maps}
We consider observations from both large-aperture telescopes (LATs) and small-aperture telescopes (SATs), which are designed to probe different angular scales and therefore have different beam resolutions, sky coverage, and noise properties. The LAT observations have relatively high angular resolution and are optimised for measuring small-scale CMB anisotropies, extending down to arcminute scales over a wide fraction of the sky. Because of this, LAT maps are particularly useful for reconstructing the CMB lensing potential and for building lensing-induced $B$-mode templates. In contrast, the SAT observations are designed to provide deeper measurements over a smaller sky region. Their main target is the large-angular-scale CMB polarisation signal, where the primordial tensor contribution to the $B$-mode polarisation is expected to be most significant. The $B$-mode polarisation measured by the SATs is therefore used to constrain the tensor-to-scalar ratio, $r$.

\begin{figure}
    \centering
    \includegraphics[width=1.\linewidth]{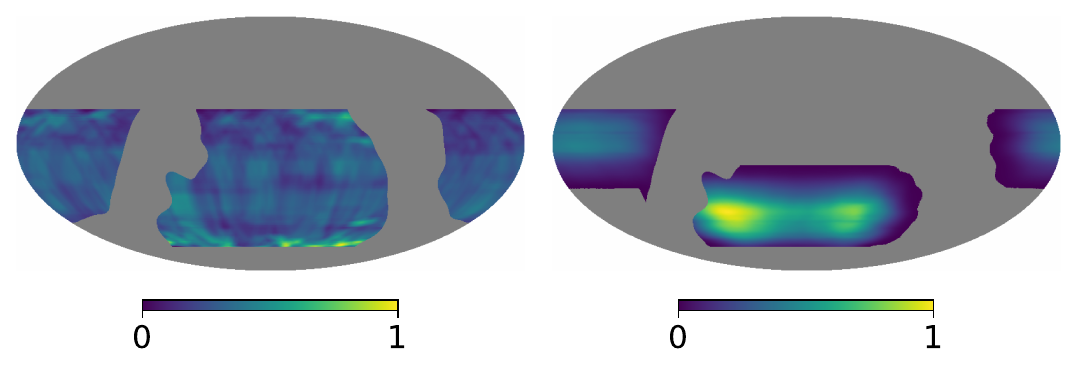}
    \caption{Normalised hit-count maps for LAT(left) and SAT(right) configurations. We multiply the hitmaps with a  binary Galactic plane mask to avoid foreground contamination. The hit-count maps are taken from \url{https://portal.nersc.gov/project/sobs/so_mapsims_data/v0.2/}. The Galactic plane mask is the Planck \texttt{GAL70} mask from Planck legacy archive.}
    \label{fig:hit_maps}
\end{figure}

\begin{figure*}
    \centering
    \includegraphics[width=0.95\linewidth]{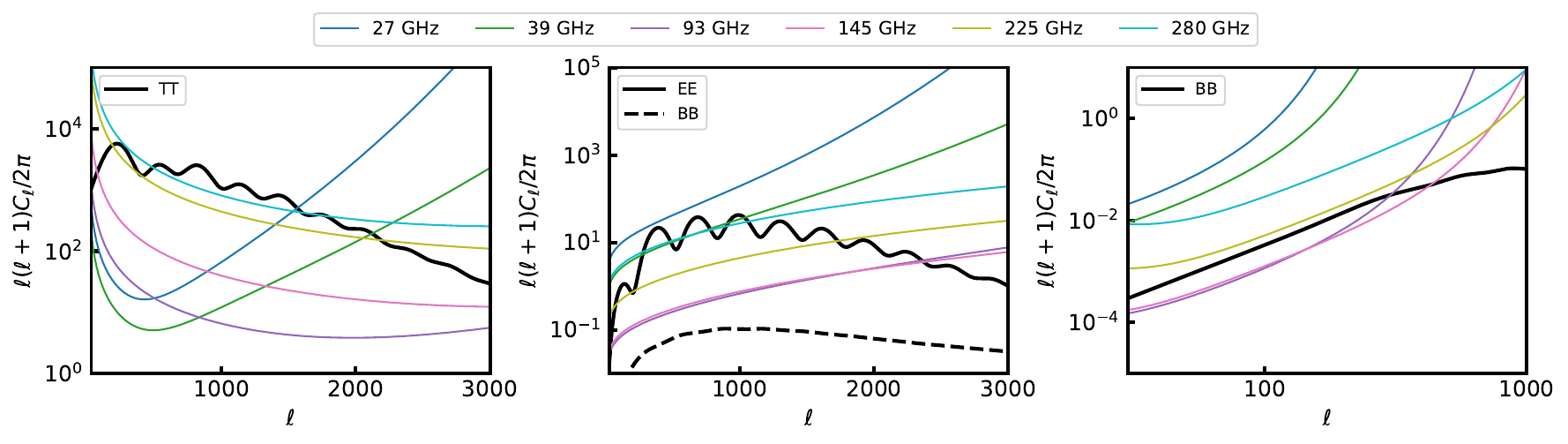}
    \caption{Beam deconvolved noise power spectra for LAT temperature (left), polarisation (middle) and SAT polarisation (right) observations shown in coloured lines for different frequency channels of SO. The solid lines are fiducial TT, EE, BB CMB spectra, and the dashed black line in middle panel is fiducial BB CMB power spectrum.}
    \label{fig:noise_ps}
\end{figure*}

\begin{figure}
    \centering
    \includegraphics[width=1.\linewidth]{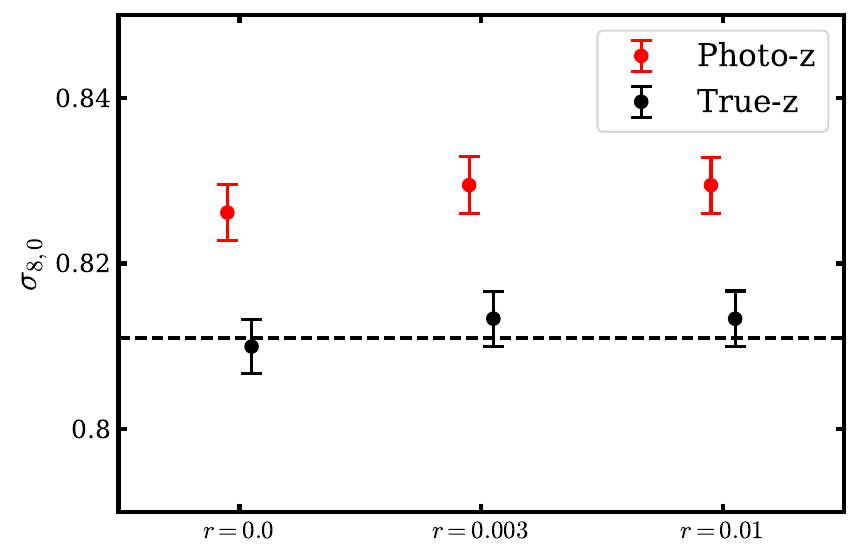}
    \caption{The estimated value of $\sigma_{8,0}$ for all three sets of simulation error bars denoting one standard deviation. The red markers show the values for inferred photometric redshift distribution and the black markers show the result for true photometric redshift distribution. The black dashed line shows the input value of $\sigma_{8,0}=0.811$}
    \label{fig:s8_estimate}
\end{figure}

The noise power spectra for both SAT and LAT frequency channels are given by \citep{SO_2019},
\begin{equation}
    N_{\ell} = N_{\rm white} + N_{\rm red} \left( \frac{\ell}{\ell_{\rm knee}}\right)^{\alpha}
\end{equation}
where, $N_{\mathrm{white}}$ is the white-noise level. The parameters $N_{\mathrm{red}}$, $\ell_{\mathrm{knee}}$, and $\alpha_{\mathrm{knee}}$ describe the amplitude, characteristic multipole, and slope of the $1/f$ noise component, respectively. Although these white-noise levels provide a good description of the noise on small angular scales, the large-scale noise is affected by low-frequency $1/f$ contributions in the detector time streams. This contamination is mainly produced by atmospheric fluctuations and electronic noise. For LAT and SAT polarisation maps, we fixed $N_{\rm red} = N_{\rm white}$. For LAT maps, we set $\ell_{\rm knee} = 1000,~\alpha_{\rm knee} = -3.4$ for temperature, and $\ell_{\rm knee}=700,~\alpha_{\rm knee}=-1.4$ for polarisation maps. The other noise and beam resolution parameters for all frequency bands of LAT and SAT are presented in Table~\ref{tab:instrument_specs}. In Fig. \ref{fig:noise_ps}, we show the LAT and SAT instrumental beam-deconvolved noise power spectra for all six frequency channels. The noise maps are generated as random Gaussian realisations of the corresponding noise power spectra. To account for the inhomogeneous depth of the survey, each noise realisation is rescaled by $1\Big/\sqrt{N^{\rm hit}_i}$, where $N^{\rm hit}_i$ denotes the hit-count map normalised such that its maximum value is set to unity. This weighting increases the noise in regions with fewer observations and reduces it in regions with higher integration time. The normalised hit-count maps used for the LAT and SAT sky coverages, corresponding to sky fractions of $f_{\rm sky}=0.4$ and $f_{\rm sky}=0.1$, respectively, are shown in Fig.~\ref{fig:hit_maps}.

We used fiducial power spectra to generate Gaussian realisation of unlensed CMB $T,~Q,~U$ maps at $N_{\text{side}}=2048$ using the Healpy package \footnote{\url{https://healpy.readthedocs.io/en/latest/}} \citep{Gorski_2005}. We produced three sets of unlensed CMB maps : Set-A, Set-B and Set-C, with input tensor-to-scalar ratio $r=0$, $r=0.003$ and $r=0.01$ respectively. For each set of CMB maps, we generated a lensing convergence map and correlated galaxy catalogue simulations (described in Section \ref{sec:lsst_simulations}). The unlensed CMB maps were then lensed with their corresponding lensing convergence map on the full sky at $N_{\text{side}}=2048$ \citep{Reinecke_2023}, using the \texttt{lenspyx}\footnote{\url{https://github.com/carronj/lenspyx}} package. We convolved the lensed CMB maps with corresponding beam of each frequency channels. The SAT maps were downgraded to $N_{\text{side}}=512$ and $\ell_{\text{max}}=1500$, while we kept the LAT maps at $N_{\text{side}}=2048$ and $\ell_{\text{max}}=6144$. We then added the noise realisation and Galactic foreground component to each frequency channel maps in LAT and SAT configurations. The Galactic foreground maps were generated as a combination of \texttt{d12} model for thermal dust and \texttt{s7} model for synchrotron emission using the \texttt{pysm}\footnote{\url{https://pysm3.readthedocs.io/en/latest/}} package \citep{Thorne_2017, Borrill_2025}. The \texttt{d12s7} model is the highest complexity foreground emission model in \texttt{pysm} package. The foreground maps were also convolved with corresponding beam before adding to the CMB maps. Finally, we applied the binary sky coverage mask for LAT and SAT configuration accordingly.

\section{Methodology}
\label{sec:methods}
\subsection{Covariance estimation}\label{sec:jackknife}
\begin{figure*}
    \begin{subfigure}[b]{0.45\textwidth}
        \centering
        \includegraphics[width=\textwidth]{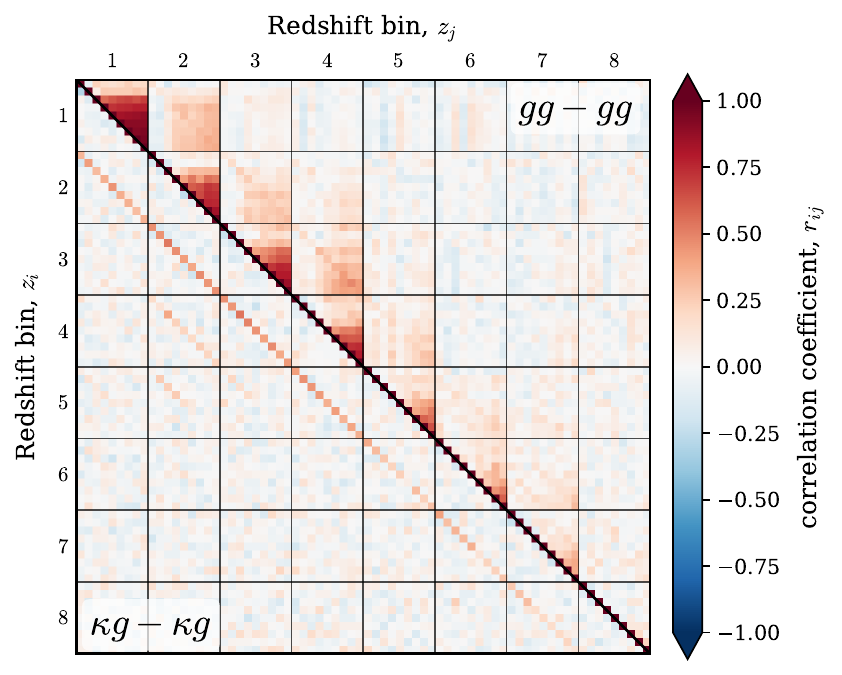}
        \caption{$gg-gg$ (uppper triangle) and $\kappa g-\kappa g$ (lower triangle) correlation matrix blocks for the eight tomographic redshift bins. Each pixel correspond to binned multipoles in the respective redshift bin.}
    \end{subfigure}
    \hfill % Horizontal space between subfigures
    % Second Subfigure
    \begin{subfigure}[b]{0.45\textwidth}
        \centering
        \includegraphics[width=\textwidth]{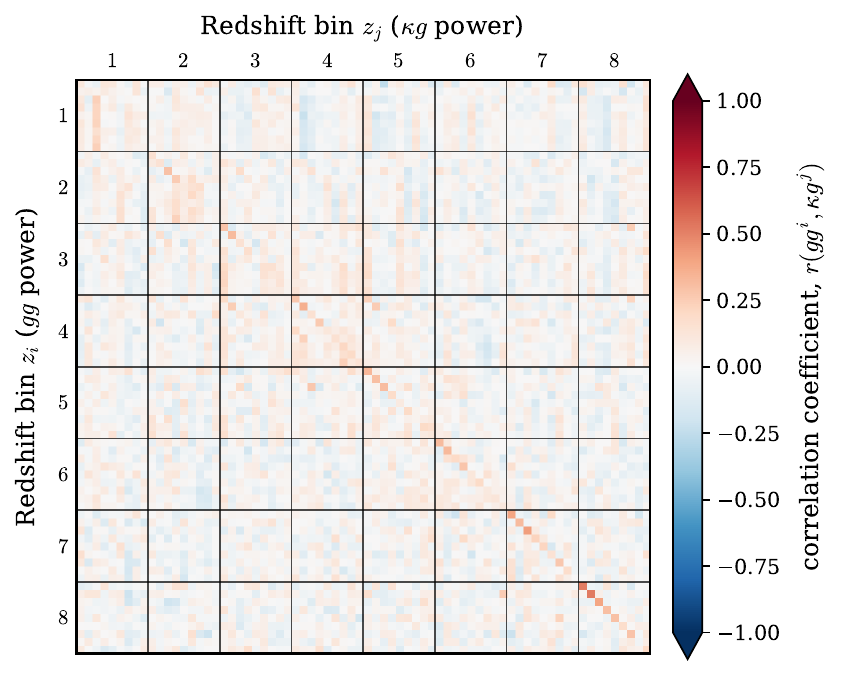}
        \caption{$\kappa g-gg$ correlation matrix blocks for the eight tomographic redshift bins. Each pixel correspond to binned multipoles in the respective redshift bin.}
    \end{subfigure}
    \caption{Correlation matrix, $r_{ij}$ as defined in Eq. \ref{eq:corr_mat}}
    \label{fig:correlation_matrix_galaxy}
\end{figure*}

We implemented the delete-one jackknife sampling to estimate the covariance matrices of galaxy-galaxy auto and lensing convergence-galaxy cross power spectra. To compute the jackknife covariance, the galaxy overdensity maps were divided into $N_{\rm JK} = 149$ disjoint sub-regions for each redshift bin and the power spectra recomputed each time by removing one jackknife region from the data. We estimated the power spectra using Namaster\footnote{\url{https://namaster.readthedocs.io/en/latest/}} package, which implements the MASTER algorithm \citep{Hivon_2002, Alonso_2019}. We used the binned spectra to calculate the covariance with the following bin boundaries - $[0, 30, 60, 90, 140, 250, 400, 600, 900, 1200, 1650, 2048]$. Hereafter, we denote the binned indices with $\ell$ (or $\ell'$). The jackknife covariance is then calculated as,
\begin{equation}
{\rm Cov}_{\ell\ell'} = \frac{N_{\rm JK}-1}{N_{\rm JK}}
\sum_{i=1}^{N_{\rm JK}} \left(\widehat{C}_{\ell}^{x,i}-\overline{C}_{\ell}^{x}\right)
\left(\widehat{C}_{\ell'}^{y,i}-\overline{C}_{\ell'}^{y}\right),
\end{equation}
where ${x,y} \in \{\kappa, g\}$ and $\widehat{C}_{\ell}^{x,i}$ is the power spectrum measured after excluding the $i$-th jackknife region, and
\begin{equation}
\overline{C}_{\ell}^{x} = \frac{1}{N_{\rm JK}}
\sum_{i=1}^{N_{\rm JK}} \widehat{C}_{\ell}^{x,i}
\end{equation}
is the mean power spectrum over all jackknife samples. The corresponding correlation matrix defined as
\begin{equation}
r_{\ell \ell'} = \frac{{\rm Cov}_{\ell \ell'}}
{\sqrt{{\rm Cov}_{\ell\ell}{\rm Cov}_{\ell' \ell'}}},  \label{eq:corr_mat}
\end{equation}
is shown in Fig. \ref{fig:correlation_matrix_galaxy}.

\subsection{Likelihood analysis}
\subsubsection{Galaxy bias estimation}
To obtain the optimal delensing of CMB $B$-modes with LSS tracers, we need to estimate the redshift dependent galaxy bias, $b(z)$. Additionally, we estimated the present time matter fluctuation parameter, $\sigma_{8,0}$, in order to study the impact of photometric redshift errors. We use a Gaussian likelihood analysis on the joint data vector $\widehat{C}_{\ell} \equiv [\widehat{C}^{gg}_{\ell}, \widehat{C}^{\kappa g}_{\ell}]$ to estimate these two parameter in each redshift bins separately \citep{saraf_2024},
\begin{equation}
\begin{split}
    -2\log \mathcal{L} & \Big((b,\sigma_{8,0}) | \widehat{C}_{\ell}\Big) \\ 
    & \approx \Big[\widehat{C}_{\ell} - C_{\ell}(b,\sigma_{8,0})\Big]^{\rm T}\Big[({\rm Cov}^{-1})_{\ell\ell'}\Big]\Big[\widehat{C}_{\ell'} - C_{\ell'}(b,\sigma_{8,0})\Big]
\end{split}
\label{eq:gal_likelihood}
\end{equation}
where, $C_{\ell}(b,\sigma_{8,0}) \equiv [C_{\ell}^{gg}(b, \sigma_{8,0}), C_{\ell}^{\kappa g}(b, \sigma_{8,0})]$ is the joint fiducial power spectra vector, and $[{\rm Cov}^{-1}_{\ell\ell'}]$ is the inverse covariance matrix computed using jackknife approach described in Sect. \ref{sec:jackknife}. First we compute the theory spectra by setting the bias parameter and $\sigma_{8,0}$ parameter to unity, and then the model power spectra are given in terms of the parameters as,
\begin{equation}
    \begin{split}
    C^{gg}_{\ell}(b,\sigma_{8,0}) & \propto b_{\rm eff}^2 \; \sigma_{8,0}^2 \;, \\
    C^{\kappa g}_{\ell}(b,\sigma_{8,0}) & \propto b_{\rm eff} \; \sigma_{8,0}^2 \; . \label{eq:gal_ps_model}
    \end{split}
\end{equation}
For each redshift bin, we estimated an effective galaxy bias parameter, $b_{\rm eff}$. 

\subsubsection{Tensor-to-scalar ratio constraint}
We derive the tensor-to-scalar ratio constraints from the delensed component separated $B$-mode maps. We adopted the Hamimeche-Lewis (H-L) likelihood approximation which fully captures the non-Gaussian statistics introduced by the cut-sky analyses and residual Galactic foregrounds \citep{Hamimeche_2008}. The H-L log likelihood for a single delensed $B$-mode power spectra,
$\widehat{C_{\ell}}$, is written as \citep{Deka_2026},
\begin{equation} 
    -2 \ln\mathcal{L}\left(r|\hat{C}_\ell\right) = \sum_{\ell \ell'} \left[ g(\hat{C}_\ell/C_\ell)\, C_{\ell} \right]
    \left[\textbf{M}^{-1}\right]_{\ell\ell'} \left[ g(\hat{C}_{\ell'}/C_{\ell'})\, C_{{\ell'}} \right],
    \label{eq:cmb_likelihood}
\end{equation}
where
\[ g(x) = \text{sign}(x-1) \sqrt{2(x-\ln(x)-1)}~, \]
and $\rm{sign}(x)$ denotes the signum function. $\textbf{M}$ is the covariance matrix from the simulated observation using jackknife method explained in Sect. \ref{sec:jackknife}.  Here the model power spectra, $C_{\ell}$ is same as in Eq. \eqref{eq:delensed_Bmode_model} with the residual Galactic foreground power being included in the noise term, $N^{BB}_{\ell}$. The observed power spectra are computed using Namaster package and we used the binned multipoles within $30 \leq \ell \leq 300$ with a bin width of $\Delta \ell = 30$.

\begin{table*}
    \centering
    \caption{The true input values of galaxy bias, $b_{\rm true}$ along with the estimated values $b_{\rm eff}$ and $\sigma_{8,0}$ in each tomographic bins.}
    \renewcommand{\arraystretch}{1.3}
    \begin{tabular}{lcccc}
    \hline\hline
    \rule{0pt}{2.6ex} parameters  & true value & Set-A (r=0) & Set-B (r=0.003) & Set-C (r=0.01) \\
    \hline
    \rule{0pt}{2.6ex} 
    $b_1~(z \in [0, 0.17))$ & 1.05 & $1.034 \pm 0.006$ & $1.022 \pm 0.006$ & $1.022 \pm 0.007$ \\
    $b_2~(z \in [0.17, 0.31))$ & 1.1 & $1.08 \pm 0.006$ & $1.074 \pm 0.006$ & $1.074 \pm 0.006$ \\
    $b_3~(z \in [0.31, 0.39))$ & 1.18 & $1.172 \pm 0.006$ & $1.166 \pm 0.006$ & $1.166 \pm 0.006$ \\
    $b_4~(z \in [0.39, 0.49))$ & 1.26 & $1.22 \pm 0.006$ & $1.215 \pm 0.006$ & $1.215 \pm 0.006$ \\
    $b_5~(z \in [0.49, 0.61))$ & 1.35 & $1.318 \pm 0.006$ & $1.312 \pm 0.006$ & $1.311 \pm 0.006$ \\
    $b_6~(z \in [0.61, 0.75))$ & 1.45 & $1.430 \pm 0.006$ & $1.422 \pm 0.006$ & $1.422 \pm 0.006$ \\
    $b_7~(z \in [0.75, 0.93))$ & 1.7 & $1.654 \pm 0.007$ & $1.646 \pm 0.007$ & $1.646 \pm 0.007$ \\
    $b_8~(z \in [0.93, 3))$ & 2.0 & $1.977 \pm 0.008$ & $1.968 \pm 0.008$ & $1.968 \pm 0.008$ \\
    $\sigma_{8,0}$ & 0.811 & $0.826 \pm 0.003$ & $0.829 \pm 0.003$ & $0.829 \pm 0.003$ \\
    \hline
\end{tabular}
    \label{tab:bias_estimates}
\end{table*}

\section{Results}
\label{sec:results}

\subsection{Galaxy bias and $\sigma_8$ estimates} \label{sec:bias_estimate}
We estimated galaxy bias assuming a constant value in each tomographic bin and the $\sigma_{8,0}$ parameter for all three simulation sets. To compute the model power spectra present for the likelihood analysis, we used the inferred photometric redshift distribution that are obtained by stacking photometric posteriors of the LSST galaxies. We used the jackknife matrix to perform the likelihood analysis using Eq.~(\ref{eq:gal_likelihood}). To validate our pipeline, we also considered an alternative spectroscopic galaxy survey with similar LSST redshift distribution where the true photometric redshift distributions of the bins are known. We implemented our pipeline to estimate the parameters for this spectroscopic galaxy dataset and we correctly recovered the input fiducial parameters. We used this spectroscopic survey dataset only as a benchmark in our subsequent delensing analysis.

\begin{figure}
    \centering
    \includegraphics[width=0.9\linewidth]{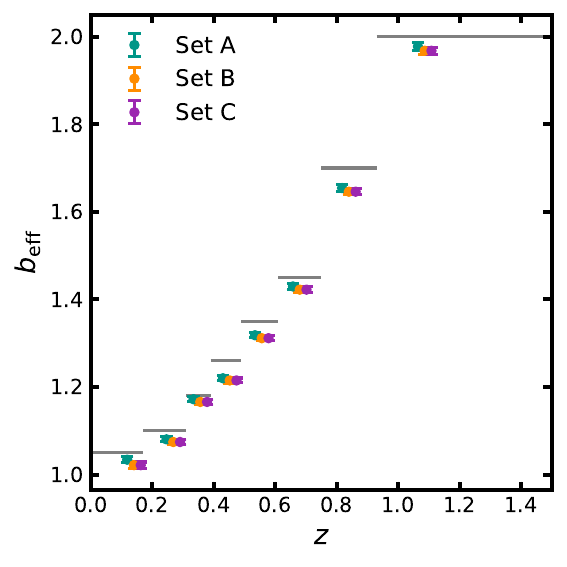}
    \caption{Recovered values of galaxy bias for each bin for three simulation sets. The horizontal line marks the true values of galaxy bias for each redshift bin and its width shows the width of the redshift bin. The $x$-axis values have been around the effective redshift for illustrative purposes (with shift value being -0.022 for Set A, 0.0 for Set B and 0.022 for Set C).}
    \label{fig:bias_estimates}
\end{figure}

  In Fig. \ref{fig:s8_estimate}, we compare the fit values of the $\sigma_{8,0}$ parameter for the three sets obtained by considering the inferred photometric redshift distribution and the true underlying photometric redshift distribution for a similar spectroscopic survey. We see the estimate of $\sigma_{8,0}$ has a bias towards higher value and  $\sigma_{8,0}$ estimate for Set-C is more than one standard deviation away from the fiducial value of $\sigma_{8,0}=0.811$.  On the hand, the galaxy bias parameters for each simulation set are lower than the input bias values, as shown in Table \ref{tab:bias_estimates} and Fig. \ref{fig:bias_estimates}. We computed the combinations of these parameters, mainly $b_{\rm eff}^2\,\sigma_{8,0}^2$ and $b_{\rm eff}\,\sigma_{8,0}^2$, which appears in the galaxy-galaxy auto power spectra and galaxy-lensing convergence cross power spectra, respectively, as shown in Eq. \eqref{eq:gal_ps_model}. These two combinations shows lower bias from the input values across all bins except the highest redshift bin ($z \in [0.93, 3.0]$) which is mainly due to redshift errors shifting the effective bias further away from its fiducial input value across this broad final bin.

This biased estimate of the parameters arises due to the fact that the inferred photometric distribution is just an approximation and it is not exactly the same as the underlying true distribution. To see the impact of this disagreement in power spectra level, we used the fitted parameter values to compute the theoretical power spectra using the inferred redshift distribution. We find agreement between the observed and theoretical galaxy-galaxy and lensing convergence-galaxy power spectra within the cosmic variance limit. The fit parameters could reproduce the observed cross-power spectra between the nearest neighbouring bins within the cosmic variance limit, except for Bin-6 and Bin-7 cross-correlation power. However, the observed cross-power spectra for the second-nearest neighbouring bins show biases of around 100\%--200\% for all bin combinations. The inferred photometric distribution fails to take into account cross bin correlations accurately.

\begin{figure}
    \centering
    \includegraphics[width=0.9\linewidth]{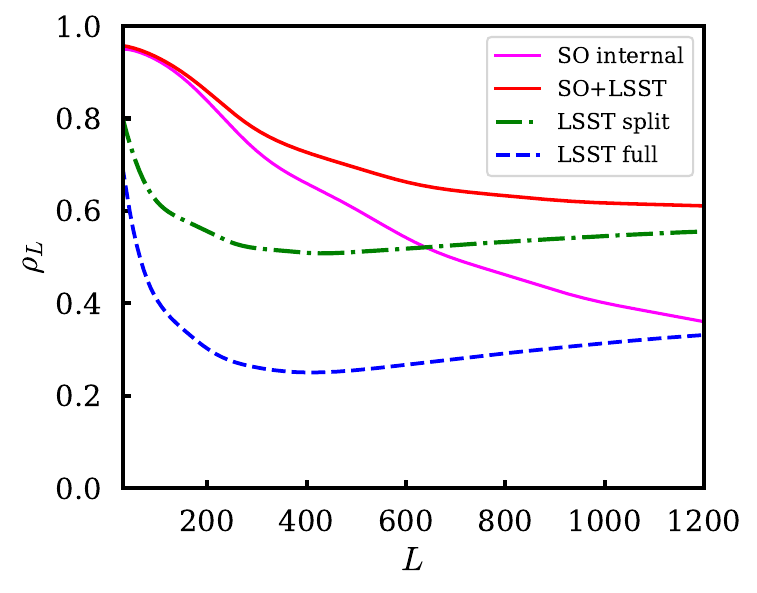}
    \caption{Correlation coefficient for different tracers and their combinations. The blue dashed line shows LSST catalogue used as a single tracer without tomographic binning. The green dashed-dot line shows the case when LSST catalogue is divided into tomographic bins. The magenta line shows the internal delensing efficiency when only CMB data is used for delensing. The red line is the optimal delensing efficiency obtained by combining CMB and LSST tomographic bins.}
    \label{fig:correlation_coeff}
\end{figure}

\begin{figure*}
    \centering
    \begin{subfigure}[t]{0.33\textwidth}
        \centering
        \includegraphics[width=0.95\linewidth]{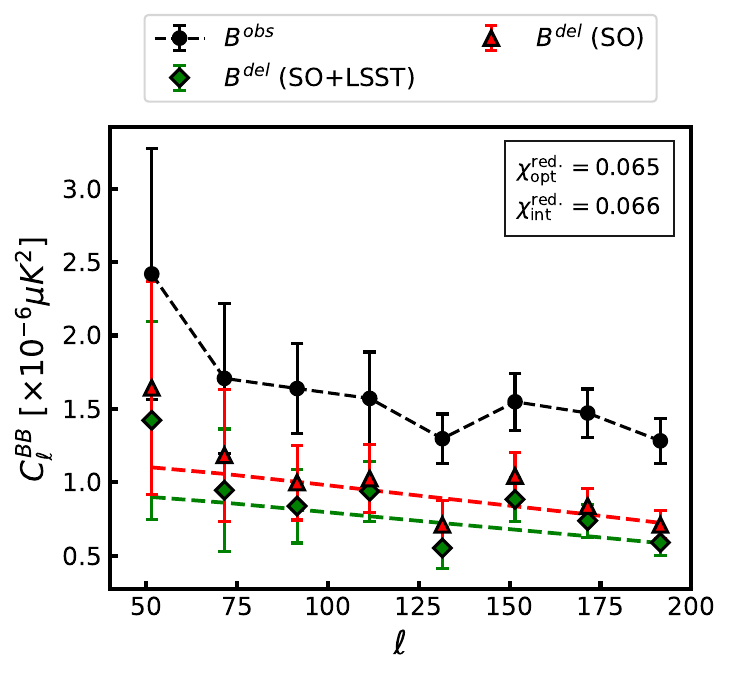}
    \end{subfigure}%
    \begin{subfigure}[t]{0.33\textwidth}
        \centering
        \includegraphics[width=0.95\linewidth]{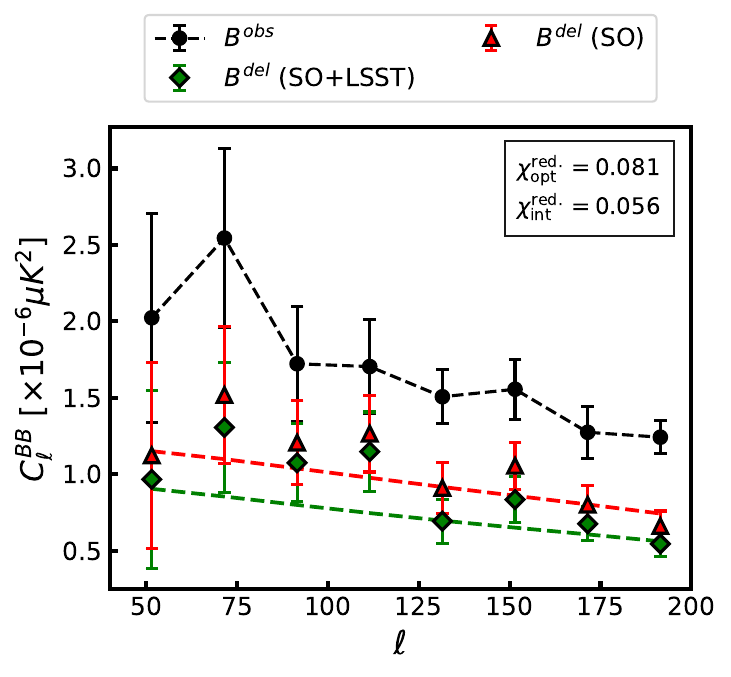}
    \end{subfigure}%
    \begin{subfigure}[t]{0.33\textwidth}
        \centering
        \includegraphics[width=0.95\linewidth]{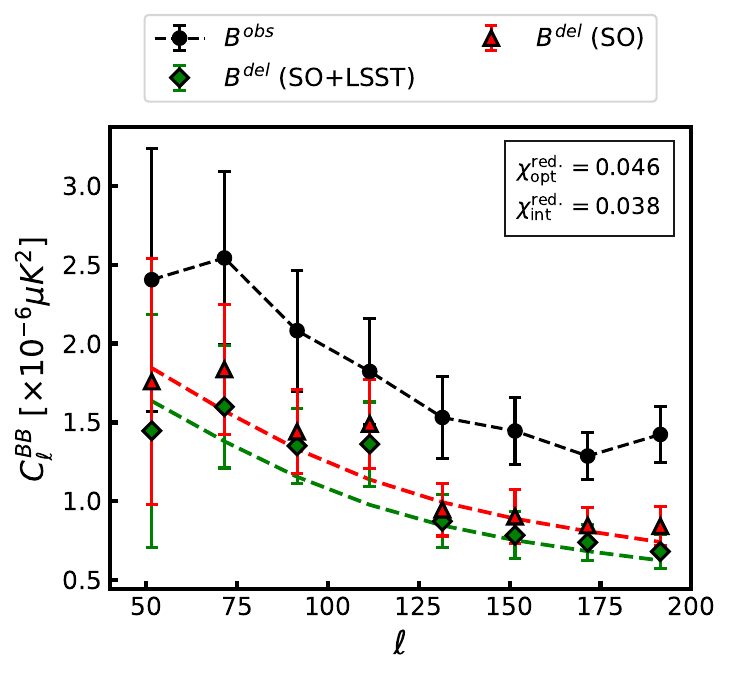}
    \end{subfigure}
    \caption{Delensed B-mode power spectra (noise subtracted) for all three sets of simulation. The black circles show the HILC cleaned SAT B-mode power spectra before delensing, and the red triangles and green diamonds show the internally delensed case and optimally delensed case, respectively. The green and red dashed lines show the fiducial power spectra for the estimated values of $r$ and $A_{\rm lens}$.}
    \label{fig:delensed_ps}
\end{figure*}

\subsection{Delensing efficiency}
We compared two methods of template based delensing in this work: an internal delensing method using only CMB observations and a multi-tracer approach combining CMB maps and galaxy catalogue. The internal delensing method uses the QE reconstructed lensing convergence field to build the lensing template using Eq. \eqref{eq:lensing_Bmode}. We computed the correlation coefficient for internal delensing case using Eq. \eqref{eq:rho_internal}. In the multi-tracer approach, we combined the CMB and galaxy overdensity maps to obtain higher delensing efficiency. We used Eq. \eqref{eq:rho_optimal} to compute the correlation coefficient using LSST galaxy maps as external tracers where we considered two cases : In one case, we split the LSST galaxy catalogue into photometric redshift bins to get multiple tomographic tracers and we refer to this case as LSST-split. In the other case, we take the LSST catalogue as a single tracer which is referred here as LSST-full. In all cases, including internal, LSST-full and LSST-split, the correlation coefficient is computed using theoretical power spectra that assumes a fiducial cosmology. As shown in Fig. \ref{fig:correlation_coeff}, we obtained higher value of correlation coefficient across all multipoles in the LSST-split case. Our results are in agreement with the work presented in \citet{Manzotti_2017} which showed that tomographic binning enhances the spatial overlap between the galaxy distribution and the lensing convergence kernel, thereby increasing their cross-correlation coefficient. 

We see that internal delensing have higher correlation at low multipoles up to $\ell < 400$, meanwhile combination LSST tomographic bins gives better correlation at high multipoles. We combined the lensing convergence estimate with tomographic LSST galaxy maps to obtain the highest level of correlation which can act as a optimal tracer for delensing. In Fig. \ref{fig:correlation_coeff} we show the optimal level of correlation coefficient obtained for combination of internal SO CMB and external LSST galaxy tracers splitted into eight redshift bins. The optimal weights to combine these tracers are computed from the theoretical power auto and cross power spectra in Eq. \eqref{eq:optimal_weights}. The theoretical power spectra are computed using the fitted bias and $\sigma_{8,0}$ parameters mentioned in Sect. \ref{sec:bias_estimate} .

In Fig. \ref{fig:delensed_ps}, we show the angular power spectra of foreground cleaned SAT B modes along with the power spectra of delensed B modes obtained from the internal delensing method and the optimal SO$+$LSST-split multi-tracer delensing method. For all three sets of simulations with different tensor-to-scalar ratio, we achieved significantly higher level of delensing performance for the case of optimal combination of SO and LSST datasets.

\subsection{$r$ and $A_{\rm lens}$ estimates}
In this section, we present the tensor-to-scalar ratio, $r$, and $A_{\rm lens}$ parameter estimated from the delensed B-mode spectra using the likelihood given by Eq. (\ref{eq:cmb_likelihood}). In the analysis, the covariance matrices were estimated using delete-one jackknife method described in Sect. \ref{sec:jackknife}. To compute the jackknife covariance matrix we divided the delensed map into 258 disjoint regions and implement a delete-one method of jackknife covariance estimation as described in Sec. \ref{sec:jackknife}. We applied the Hartlap factor correction to the estimation of the inverse covariance matrices to avoid any bias \citep{Hartlap_2007}. For the likelihood joint-analysis of $r$ and $A_{\rm lens}$, we used uniform priors for all three simulation sets : $r^{\rm prior} \subset \mathcal{U}[0,0.03]$ and $A_{\rm lens}^{\rm prior} \subset \mathcal{U}[0,1]$. Here, $\mathcal{U}[x,y]$ denotes a uniform distribution that has unity value within the interval $[x,y]$ and zero elsewhere. In Table \ref{tab:estimate_r_alens}, we present the derived values of $r$ and $A_{\rm lens}$ for our simulation sets. We also ran the likelihood analysis for different priors on $r$, for example, a half-normal prior with mean at $r=0$, a log normal prior and a Gaussian prior.  We use the uniform prior results as a our baseline for all three simulation sets.

\begin{figure}
    \centering
    \includegraphics[width=1.\linewidth]{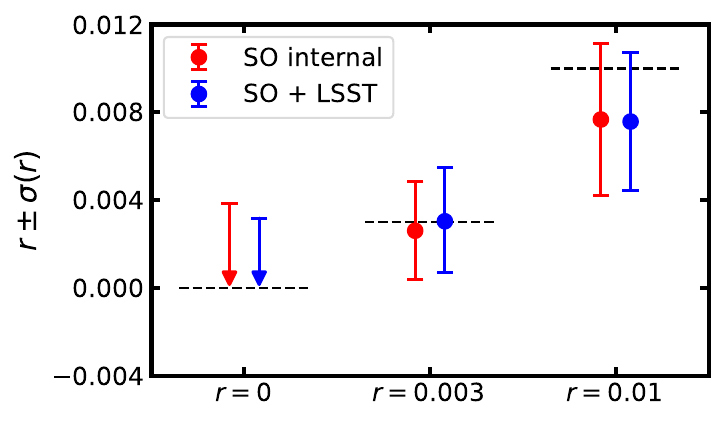}
    \caption{Estimated value of $r$ from all simulation sets for SO internal delensing (shown in red) and SO+LSST optimal delensing (shown in blue). The dashed black line shown the input value of $r$.}
    \label{fig:estimated_r}
\end{figure}

\begin{table*}
    \centering
    \caption{The estimated and true values $r$ and $A_{\rm lens}$ for all three simulation sets and for the two different delensing method. The prior used for $r$ and $A_{\rm lens}$ are uniform : $r^{\rm prior} \subset \mathcal{U}[0,0.03]$ and $A_{\rm lens}^{\rm prior}  \subset \mathcal{U}[0,1]$.}
    \renewcommand{\arraystretch}{1.3}
    \begin{tabular}{cccccc}
    \hline\hline
    \rule{0pt}{2.6ex} Parameter & Set  & True value & Prior & SO internal & SO+LSST optimal \\
    \hline
    \rule{0pt}{2.6ex} 
       & Set-A & 0 & $\mathcal{U}[0,0.03]$ & $< 0.0038$ & $< 0.0032$  \\
    $r$ & Set-B & 0.003 & $\mathcal{U}[0,0.03]$ & $0.0026 \pm 0.0022$ & $0.003^{+0.0023}_{-0.0025}$ \\
       & Set-C & 0.01 & $\mathcal{U}[0,0.03]$ & $0.0077 \pm 0.0035$ & $0.0076 \pm 0.0031$ \\ \hline
    \rule{0pt}{2.6ex}  
      & Set-A & -- & $\mathcal{U}[0,1]$ & $0.53 \pm 0.05$ & $0.43 \pm 0.05$  \\
   $A_{\rm lens}$ & Set-B & -- & $\mathcal{U}[0,1]$ & $0.55 \pm 0.04$ & $0.41 \pm 0.04$ \\
       & Set-C & -- & $\mathcal{U}[0,1]$ & $0.54 \pm 0.05$ & $0.46 \pm 0.05$ \\
    \hline
\end{tabular}
    \label{tab:estimate_r_alens}
\end{table*}

In Fig. \ref{fig:estimated_r} we show the estimated value of $r$ and its uncertainty for internal and optimal delensing methods. We obtained the estimated value of $r$ as the mode of the posterior distribution. We then show the high density interval of 68\%  as the upper and lower bound on the $r$ estimates. For the input $r=0.$ case, the lower bound is set to be $r=0$. As shown in Fig. \ref{fig:estimated_r}, we are able to recover the input value of $r$ within the statistical uncertainties in all cases. For $r=0.01$, we obtained a $2\sigma$ detection of our input tensor B-mode signal. Although the estimated parameter value for $r=0.01$ remains consistent with the true input value within one standard deviation, a slight  underestimation is observed.

\begin{figure}
    \centering
    \includegraphics[width=1.\linewidth]{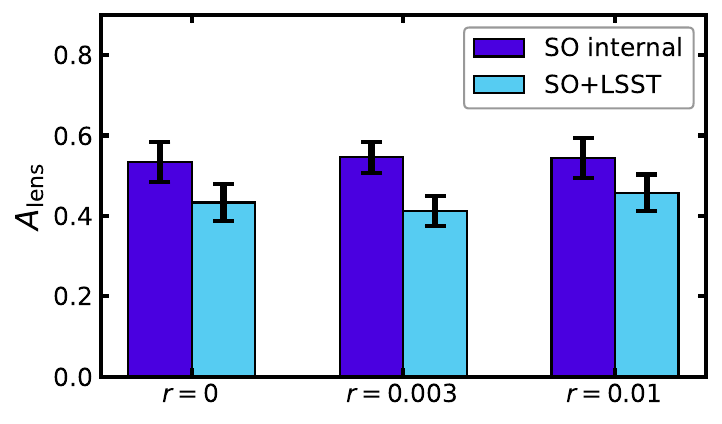}
    \caption{Estimated value of $A_{\rm lens}$ for all three simulation sets. The error bar shows $1\sigma$ error on the value. The deep blue bars are for the SO internal delensing and the sky blue bars are for the SO+LSST optimal delensing case.}
    \label{fig:estimated_alens}
\end{figure}

Figure~\ref{fig:estimated_alens} presents the reconstructed $A_{\rm lens}$ values for all three simulation set. From Fig. \ref{fig:estimated_alens}, we see that, joint optimal delensing outperforms SO internal delensing across all three sets, reducing the residual lensing power $A_{\rm lens}$ by 10–20\%. This clearly demonstrates the enhanced delensing efficiency gained by combining tomographic LSST galaxy maps with SO CMB observations.

\section{Summary}
\label{sec:summary}
In this work, we explored the implementation of multi-tracer delensing approach with upcoming CMB datasets from SO and LSST Y1 observations. We used realistic simulations of SO and LSST observations that include Galactic foreground contamination in CMB maps and photometric redshift errors in LSST measurements. We used a high complexity Galactic foreground emission model in the CMB maps which contains non-Gaussian small scales to mimic real-world observations. We used a harmonic internal linear combination (HILC) method for component separation which is a non-parametric method with no prior parameterisation of the foreground emissions. With our pipeline, we generated three sets of simulations of CMB maps and lensing convergence along with correlated galaxy catalogues from input fiducial power spectra. These three sets had three different input values of $r$ while other cosmological parameters were kept fixed. 

The photometric redshift of each galaxies were sampled to have a photometric redshift error expected for LSST systematics. We divided our catalogue in tomographic bins to obtain multiple tracers across the redshift range coverage of LSST which we later used in our multi-tracer delensing pipeline. First, we estimated the galaxy bias and matter fluctuation amplitude for each tomographic redshift bin from the measured galaxy-galaxy auto and galaxy-CMB lensing convergence cross power spectra. Due to the presence of photometric redshift errors we observed systematic biases in the recovered parameters. We then generated theoretical power spectra from the best-fit $b_{\rm eff}$ and $\sigma_{8,0}$ parameters and computed the optimal weights to combine the multiple tracers for delensing of CMB $B$-modes. We compared the delensing efficiency between internal CMB delensing method and an optimal delensing with combinations of CMB lensing convergence and galaxy overdensity maps. We showed that optimally combined tomographic redshift bins of LSST galaxy overdensities with CMB maps improved the delensing efficiency by 10-20\% compared to using CMB internal delensing alone. With this optimal delensing approach, we achieved a lensing residue value of around $A_{\rm lens}$=0.4. We do not see any bias in the estimate of the tensor-to-scalar ratio due to the presence of photometric redshift errors and we are able to recover fiducial values of $r$ in all three cases, within $\pm 1\,\sigma$ uncertainties. 

It is worth noticing that \citet{Hertig_2024} reported a better delensing performance ($A_{\rm lens} = 0.35$), driven predominantly by the inclusion of CIB as an extra tracer field and a wider redshift distribution of galaxies with a median peak at $z\sim1$ which has a larger fraction of overlap with the lensing kernel compared to the redshift distribution used in our work. Our estimated uncertainties on $r$ are approximately twice as large in comparison to \citet{Hertig_2024}. However, our analysis incorporates a more realistic photometric redshift distribution alongside highly non-Gaussian Galactic foreground templates from \texttt{PySM}.

There are several improvements possible over the current work to refine the constraints on $r$ as described below:.
\begin{enumerate}
    \item The photometric redshift errors do not produce any bias in the tensor-to-scale ratio, but shifts $\sigma_{8}$ to higher values. In this work, we stacked the posteriors of galaxies photometric redshifts to emulate the redshift distribution in tomographic bins. A proper treatment of the impact of photometric redshift errors on the redshift distribution of tomographic bins, such as clustering redshifts or including stretch-shift parameters, is necessary to obtain unbiased estimates of $\sigma_{8}$ and galaxy bias.
    \item Although we used \texttt{RAIL}, to sample the LSST galaxies in our simulations, the resulting redshift distribution of galaxies is shallow, with $z_{\text{median}}\sim 0.5$, and does not fully represent the expected distribution of galaxies to be observed by the Rubin Observatory (expected $z_{\text{median}}\sim 0.75$ for Y1). This is because, \texttt{RAIL} utilises the DES-calibrated Buzzard $N$-body simulations \citep{Derose_2019}, whose galaxy population and photometric limits reflect a shallower baseline. A deeper redshift distribution of galaxies will optimise the delensing of $B$-modes and tighten the constrain on $r$.
    \item Including other tracers such as CIB, and radio continuum galaxies and 21-cm intensity mapping from the Square Kilometre Array (SKA) Observatory will help increase the delensing efficiency to up to 70\%.
\end{enumerate}
A systematic study including these improvements falls out of the scope of the work presented in this paper, where we studied the constraints on tensor-to-scalar ratio with near-realistic simulations of SO and LSST. We reserve the described extensions for future work to investigate the role of observational systematics on constraints of tensor-to-scalar ratio.

%%%%%%%%%%%%%%%%%%%%%%%%%%%%%%%%%%%%%%%%%%%%%%%%%%%%%%%%%%%%%%
\begin{acknowledgements}
KD acknowledges partial support from STER programme BPI/STE/2021/1/00033/U/00001 funded by NAWA. KD acknowledges Cosmoverse COST action STSM grant (CA21136) for travel support. KD was partially supported by the Polish National Agency for Academic Exchange under the CLEVER, Strategic Partnership programme BPI/PST/2024/1/00019. CSS is supported by the KASI project `Research on the principles of the accelerating expansion of the Universe' (grant no. 2026183201). The authors acknowledge the use of the public software packages : CAMB, healpy, lenspyx, cmblensplus, plancklens, NaMaster, emcee, PySM3, Numpy, Scipy and matplotlib. The authors are thankful for the computational resources provided by the CI\'{S} NCBJ cluster. This work used the NASA Astrophysics Data System Bibliographic Services.
\end{acknowledgements}

\bibliographystyle{aa} % style aa.bst
\bibliography{refer} % your references Yourfile.bib

\end{document}